\documentclass[%
 reprint,
superscriptaddress,
showpacs,preprintnumbers,
 amsmath,amssymb,
 aps,
prc,
]{revtex4-1}

\usepackage{graphicx}% Include figure files
\usepackage{dcolumn}% Align table columns on decimal point
\usepackage{bm}% bold math
\usepackage{epstopdf}
\usepackage{multirow}
\usepackage[inline]{enumitem}
\usepackage{paralist}
\usepackage[version=3]{mhchem}

\begin{document}

\preprint{APS/123-QED}

\title{Laser spectroscopy of francium isotopes at the borders of the region of reflection asymmetry}% Force line breaks with \\

\author{I. Budin\v{c}evi\'{c}}
\email{ivan.budincevic@fys.kuleuven.be}
\affiliation{KU Leuven, Instituut voor Kern- en Stralingsfysica, B-3001 Leuven, Belgium}

\author{J. Billowes}
\affiliation{School of Physics and Astronomy, The University of Manchester, Manchester M13 9PL, UK}
\author{M. L. Bissell}
\affiliation{KU Leuven, Instituut voor Kern- en Stralingsfysica, B-3001 Leuven, Belgium}
\author{T. E. Cocolios}
\affiliation{School of Physics and Astronomy, The University of Manchester, Manchester M13 9PL, UK}
\author{R. P. de Groote}
\affiliation{KU Leuven, Instituut voor Kern- en Stralingsfysica, B-3001 Leuven, Belgium}
\author{S. De Schepper}
\affiliation{KU Leuven, Instituut voor Kern- en Stralingsfysica, B-3001 Leuven, Belgium}
\author{V. N. Fedosseev}
\affiliation{Engineering Department, CERN, CH-1211 Geneva 23, Switzerland}
\author{K. T. Flanagan}
\affiliation{School of Physics and Astronomy, The University of Manchester, Manchester M13 9PL, UK}
\author{S. Franchoo}
\affiliation{Institut de Physique Nucl\'{e}aire d'Orsay, F-91406 Orsay, France}
\author{R. F. Garcia Ruiz}
\affiliation{KU Leuven, Instituut voor Kern- en Stralingsfysica, B-3001 Leuven, Belgium}
\author{H. Heylen}
\affiliation{KU Leuven, Instituut voor Kern- en Stralingsfysica, B-3001 Leuven, Belgium}
\author{K. M. Lynch}
\affiliation{KU Leuven, Instituut voor Kern- en Stralingsfysica, B-3001 Leuven, Belgium}
\affiliation{School of Physics and Astronomy, The University of Manchester, Manchester M13 9PL, UK}
\affiliation{Physics Department, CERN, CH-1211 Geneva 23, Switzerland}

\author{B. A. Marsh}
\affiliation{Engineering Department, CERN, CH-1211 Geneva 23, Switzerland}
%\author{P. J. R. Mason}
%\affiliation{Department of Physics, University of Surrey, Guildford, GU2 7XH, UK}
\author{G. Neyens}
\affiliation{KU Leuven, Instituut voor Kern- en Stralingsfysica, B-3001 Leuven, Belgium}
\author{T. J. Procter}
\thanks{Present address: TRIUMF, Vancouver, British Columbia V6T
2A3, Canada.}
\affiliation{School of Physics and Astronomy, The University of Manchester, Manchester M13 9PL, UK}
\author{R. E. Rossel}
\affiliation{Engineering Department, CERN, CH-1211 Geneva 23, Switzerland}
\affiliation{Institut f\"{u}r Physik, Johannes Gutenberg-Universit\"{a}t, D-55128 Mainz, Germany}
\author{S. Rothe}
\affiliation{Engineering Department, CERN, CH-1211 Geneva 23, Switzerland}
%\affiliation{Institut f\"{u}r Physik, Johannes Gutenberg-Universit\"{a}t, D-55128 Mainz, Germany}
%\author{G. S. Simpson}
%\affiliation{Institut Laue-Langevin, B.P. 156, F-38042 Grenoble Cedex, France}
%\author{A. J. Smith}
%\affiliation{School of Physics and Astronomy, The University of Manchester, Manchester M13 9PL, UK}
\author{I. Strashnov}
\affiliation{School of Physics and Astronomy, The University of Manchester, Manchester M13 9PL, UK}
\author{H. H. Stroke}
\affiliation{Department of Physics, New York University, New York, New York 10003, USA}
%\author{P. M. Walker}
%\affiliation{Department of Physics, University of Surrey, Guildford, GU2 7XH, UK}
\author{K. D. A. Wendt}
\affiliation{Institut f\"{u}r Physik, Johannes Gutenberg-Universit\"{a}t, D-55128 Mainz, Germany}
%\author{R. T. Wood}
%\affiliation{Department of Physics, University of Surrey, Guildford, GU2 7XH, UK}

%\affiliation{%


%\affiliation{
%\affiliation{
%\affiliation{%


\date{\today}% It is always \today, today,

\begin{abstract}
%\begin{description}
\begin{inparadesc}
The magnetic dipole moments and changes in mean-square charge radii of the neutron-rich $^{218m,219,229,231}\text{Fr}$ isotopes were measured with the newly-installed Collinear Resonance Ionization Spectroscopy (CRIS) beam line at ISOLDE, CERN, probing the $7s~^{2}S_{1/2}$ to $8p~^{2}P_{3/2}$ atomic transition. The $\delta\langle r^{2}\rangle^{A,221}$ values for $^{218m,219}\text{Fr}$ and $^{229,231}\text{Fr}$ follow the observed increasing slope of the charge radii beyond $N~=~126$. The charge radii odd-even staggering in this neutron-rich region is discussed, showing that $^{220}\text{Fr}$ has a weakly inverted odd-even staggering while $^{228}\text{Fr}$ has normal staggering. This suggests that both isotopes reside at the borders of a region of inverted staggering, which has been associated with reflection-asymmetric shapes. The $g(^{219}\text{Fr}) = +0.69(1)$ value supports a $\pi 1h_{9/2}$ shell model configuration for the ground state.  The $g(^{229,231}\text{Fr})$ values support the tentative $I^{\pi}(^{229,231}\text{Fr}) = (1/2^{+})$ spin, and point to a $\pi s_{1/2}^{-1}$ intruder ground state configuration.

\end{inparadesc}
%\end{description}
\end{abstract}

\maketitle

%\tableofcontents

\section{\label{sec:Introduction}Introduction}

Nuclei possessing reflection-asymmetric shapes have attracted much theoretical and experimental attention (\cite{Butler1996} and references therein), even reaching the wider scientific community \cite{Gaffney2013}. These nuclei are located in a narrow region of the nuclear chart centered approximately around $^{225}_{89}\text{Ac}$ \cite{Sheline1987}. The neutron-rich \ce{_{87}^{220--228}Fr} isotopes located in this vicinity have already been studied with laser spectroscopy \cite{Coc1985,Duong1987} and decay spectroscopy \cite{Akovali1992,Browne2001,Aiche1988,Sheline1994,Debray2000,Liang1991,Liang1992,Liang1990,Kurcewicz1992,Burke1997,Kurcewicz1997,Fraile2001}, but there has been no clear agreement on the presence of stable octupole deformations in these nuclei. For example the ground-state spin-parities of the odd-odd \ce{^{220--228}Fr} isotopes, measured by collinear laser spectroscopy and magnetic resonance \cite{Coc1985}, were reproduced by Sheline et al. \cite{Sheline1988} using a folded Yukawa octupole deformed model \cite{Leander1984} with an octupole deformation parameter value $\epsilon_{3}= +0.08$. However, Ekstr\"{o}m et al. \cite{Ekstrom1986} reproduced the experimental spin values and qualitatively reproduced the magnetic dipole and spectroscopic quadrupole moment values for $^{224,226,228}\text{Fr}$, using the core-quasiparticle model \cite{Larsson1978}, including only quadrupole and hexadecapole deformations. Many of the $A\geq 213$ francium isotopes exhibit certain experimental signs of reflection-asymmetric nuclear shapes \cite{Butler1996}, such as the presence of parity doublet decay bands connected by enhanced E1 transitions, as in $^{217-223,225,227}\text{Fr}$\cite{Aiche1988,Debray2000,Liang1991,Liang1992,Kvasil1992,Kurcewicz1992,Burke1997,Kurcewicz1997}. Another effect that has been related to the presence of reflection-asymmetric shapes is the inversion of the mean-square charge radius odd-even staggering order, as seen in the francium, radium and radon isotopes around  $ N = 136$ \cite{Ahmad1988,Coc1987,Borchers1987}. This effect has been described by Otten \cite{Far1989}, interpreting it as corroborating the calculations by Leander and Sheline \cite{Leander1984}, which suggest that octupole deformations should be more pronounced in odd than in even nuclei. Otten also notes that the calculations by Talmi \cite{Talmi1984} imply a regular odd-even staggering for even multipole deformations and inverted odd-even staggering for odd multipole deformations. 

In this article, we report on experimental results from collinear laser spectroscopy performed for the first time on the isotopes $^{218m,219,229,231}\text{Fr}$, at the borders of the region of reflection asymmetry, using the new Collinear Resonance Ionization Spectroscopy (CRIS) beamline \cite{Flanagan2013,Coco2013}.  Measurements of the neutron-deficient isotopes of francium, down to $^{202}\text{Fr}$, were performed during the same experiment and have recently been published \cite{Flanagan2013,LynchNDef2013}. In the previous collinear laser spectroscopy study \cite{Coc1985}, the $^{218m,219}\text{Fr}$ isotopes could not be studied due to their low production yield and short half-lives $t_{1/2}(^{218m}\text{Fr}) = 22.0(5)~\text{ms}$ \cite{Ewan1982}, $t_{1/2}(^{219}\text{Fr}) = 20(2)~\text{ms}$ \cite{Meinke1951}.

\section{\label{Experiment}Experimental setup }
\begin{figure*}
\includegraphics[scale=0.6]{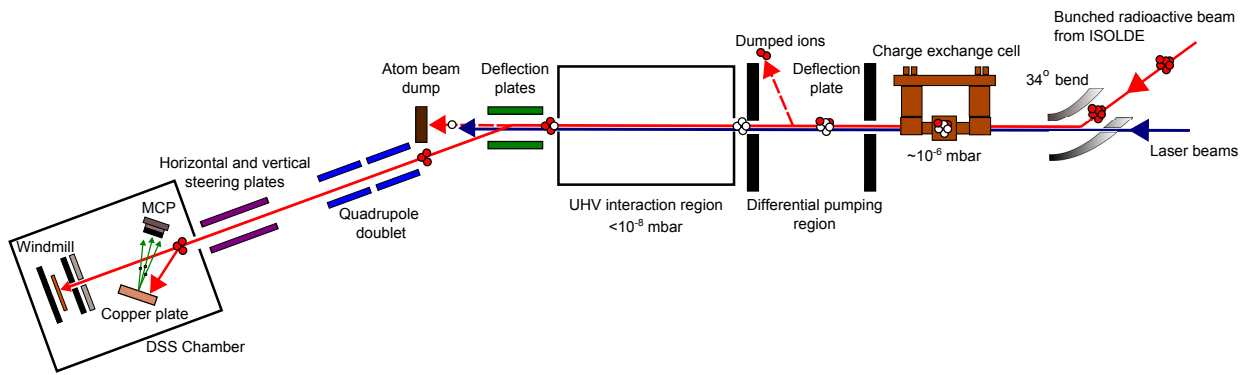}% Here is how to import EPS art
\caption{\label{fig:beam line} (Color on-line) Schematic drawing of the CRIS beam line.}
\end{figure*}

The francium isotopes of interest were produced at the ISOLDE facility in CERN, by impinging 1.4 GeV protons on a thick UC$_{x}$ target. These collisions  produced the radioactive atoms of interest via spallation. The atoms diffused out of the target to a thin transfer tube, heated to $\sim~2400^{\circ}\text{C}$ to facilitate diffusion. Surface-ionized francium ions were then accelerated to $50~\text{keV}$ and mass separated by the HRS high resolution mass separator, before being sent to the ISCOOL gas-filled segmented linear Paul trap \cite{Franberg2008,Mane2009}. The ability to bunch the beam overcomes losses associated with the duty cycle of the pulsed lasers, which previously reduced the effectiveness of the method \cite{Schulz1991}. Using ISCOOL the ions can be bunched and the time of their release from the trap synchronized with the laser-system duty cycle, thus increasing the experimental efficiency by several orders of magnitude (a production-to-detection efficiency of 1\% was measured for $^{202}\text{Fr}$ \cite{Flanagan2013}). The ion bunches leaving ISCOOL were reaccelerated to $50~\text{keV}$ and deflected to the CRIS beam line, schematically shown in Fig.~\ref{fig:beam line}.

The ion bunch was passed through a charge exchange cell filled with potassium vapour  held at $150~^{\circ}\text{C}$. During the experimental run, the potassium vapor produced a background pressure of $10^{-6}~\text{mbar}$ in the charge exchange cell region, while neutralizing the ion bunches via collisional charge exchange with a neutralization efficiency of 50 \%. Deflection plates were placed after the charge exchange cell for deflection of non-neutralized ions. After the charge exchange cell, the atomic bunch passed through a differential pumping section before arriving at the interaction region where it  interacts with the laser light. The ionization scheme used the $7s~^{2}S_{1/2} - 8p~^{2}P_{3/2}$ transition at $23658.306~\text{cm}^{-1}$($422.7~\text{nm}$) to measure the hyperfine structure and a 1064-nm nonresonant step to excite the atoms beyond the ionization threshold. The 422.7-nm transition was excited by frequency-doubled laser light from the RILIS $10~\text{kHz}$ Ti:Sa laser \cite{Rothe2011}, transported to the CRIS beam line via a 35-m optical fibre. The laser had a tunable frequency range of $\pm100~\text{GHz}$ for scanning over the hyperfine structure, with a frequency stability of $<100 \text{MHz}$ and a linewidth of around $1.5~\text{GHz}$. The $1064~\text{nm}$ laser light was produced by a $30~\text{Hz}$ Nd:YAG laser (Spectra-Physics Quanta-ray) situated near the CRIS beam line. The two laser pulses and the ion bunch from ISCOOL were synchronized to overlap in the interaction region. This was done using a Quantum Composers digital delay generator (Model: QC9258), with the 422.7-nm laser pulse serving as the master trigger. The pressure within the interaction region was kept bellow $8\times10^{-9}~\text{mbar}$ to keep the background (originating from collisional ionization) to a minimum. The experimentally observed collisional reionization efficiency was on the order of $0.001~\%$. After resonance ionization of the atomic bunch, the ions were deflected to a biased copper plate ($-600~\text{V}$), where secondary electrons were emitted upon the ion impact and guided via an electrostatic-field gradient to a micro-channel plate (MCP) detector \cite{Rajabali2013}. A LeCroy WavePro 725 Zi $2.5~\text{GHz}$ bandwidth oscilloscope was used to detect the ion signal with a time window around the ion arrival time of $10~\mu\text{s}$. During the experiment $\alpha$-decay energy spectra were collected in separate measurements by implanting the ions in a carbon foil mounted in the decay spectroscopy station (DSS), around which a dedicated $\alpha$-decay spectroscopy setup \cite{Lynch2012,LynchNDef2013,Rajabali2013} detects emitted $\alpha$-particles.

\section{\label{sec:Results}Results}

\begin{figure}
\includegraphics[width=\linewidth]{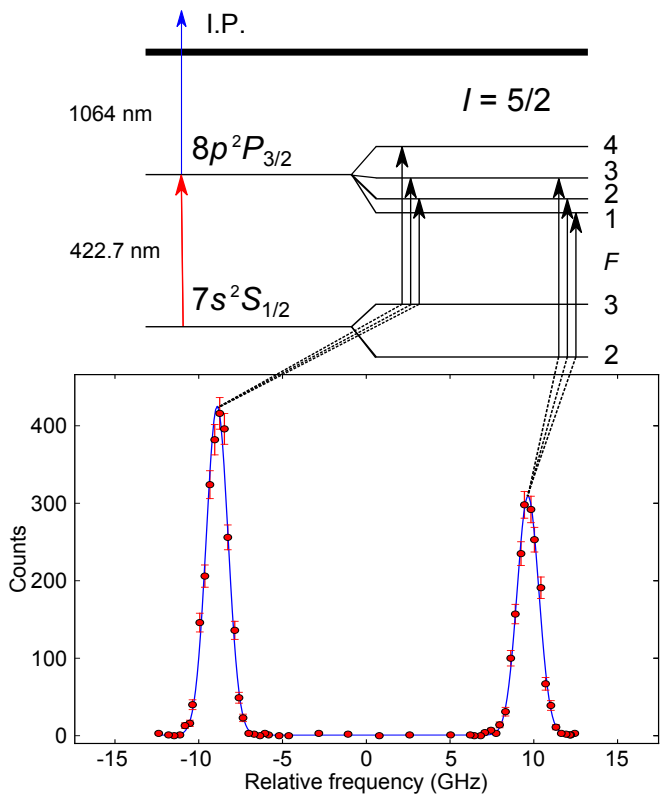}% Here is how to import EPS art
\caption{\label{fig:221_spectrum} (Color on-line)  Example hyperfine spectrum of $^{221}\text{Fr}$. The $8p~^{2}P_{3/2}$ splitting is unresolved and the $7s~^{2}S_{1/2}$ splitting is extracted from the distance between the two peaks according to equation \ref{eq:energy}.}
\end{figure}

Information about the magnetic dipole moment and changes in mean-square charge radii was obtained from the hyperfine spectra. From initial fitting of the hyperfine structure peaks with Voigt profiles, it was established that the resonance line shapes were fully dominated by a Gaussian profile, originating from the linewidth of the $422.7~\text{nm}$ RILIS Ti:Sa laser. The hyperfine structure resonance spectra were fit with a $\chi^{2}-\text{minimization routine}$, with the same full-width at half maximum for all peaks, free intensity ratios and their positions related by the hyperfine splitting energy

\begin{equation}
E_{F}=\frac{1}{2}A(F(F+1)-I(I+1)-J(J+1)),
\label{eq:energy}
\end{equation}

\noindent where $|I-J|\leq F \leq I+J$. The $A$ value is related to the magnetic dipole moment by $A=\mu_{I}B_{e}(0)/IJ$, where $B_{e}(0)$ is the magnetic field of the atomic electrons at the site of the nucleus. Due to the linewidth of $1.5~\text{GHz}$ in this initial experimental campaign only being able to resolve the lower-state splitting, the data did not provide any information on the nuclear spin. Therefore the spin $I$ has been taken from literature assignments based on decay-spectroscopy data \cite{Sheline1999,Ewan1982,Browne2001,Borge1992,Fraile2001}, and the related $A$ values extracted from the data. The upper state splitting $A(8p~^{2}P_{3/2})$ was not resolved, as illustrated for $^{221}\text{Fr}$ in Fig.~\ref{fig:221_spectrum}. The ratio $A(8p~^{2}P_{3/2})/A(7s~^{2}S_{1/2})= +0.0036$ for $^{221}\text{Fr}$ \cite{Duong1987}, was kept constant for all isotopes, neglecting the hyperfine anomaly staggering which is less than 1~\% of this ratio \cite{Grossman1999}. The magnetic dipole moments of $^{218m,219,229,231}\text{Fr}$ were evaluated relative to $^{210}\text{Fr}$:  $A_{\text{ref}}(7s~^{2}S_{1/2})(^{210}\text{Fr}) = +7195.1(4)~\text{MHz}$ \cite{Coc1985}, $I^{\pi}_{\text{ref}}(^{210}\text{Fr}) = 6^{+}$ \cite{Ekstrom1978} , and $\mu_{\text{ref}}(^{210}\text{Fr}) = +4.38(5)~\mu_{N}$ \cite{Gomez2008}, using the formula 

\begin{equation}
\mu_{\text{exp}}=\frac{A_{\text{exp}}(7s~^{2}S_{1/2})I_{\text{exp}}\mu_{\text{ref}}}{A_{\text{ref}}(7s~^{2}S_{1/2})I_{\text{ref}}}.
\end{equation}

\begin{figure}
\includegraphics[width=\linewidth]{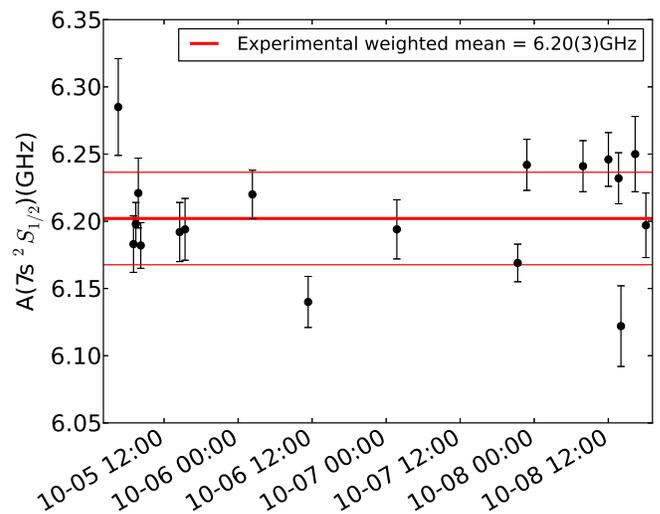}% Here is how to import EPS art
\caption{\label{fig:A_scatter} (Color on-line)  Plot showing the scatter of the $A(7s~^{2}S_{1/2})$ values extracted from 18 hyperfine spectra of the reference $^{221}\text{Fr}$ isotope taken throughout the experiment.}
\end{figure}

\begin{figure}
\includegraphics[ height = 0.8 \linewidth]{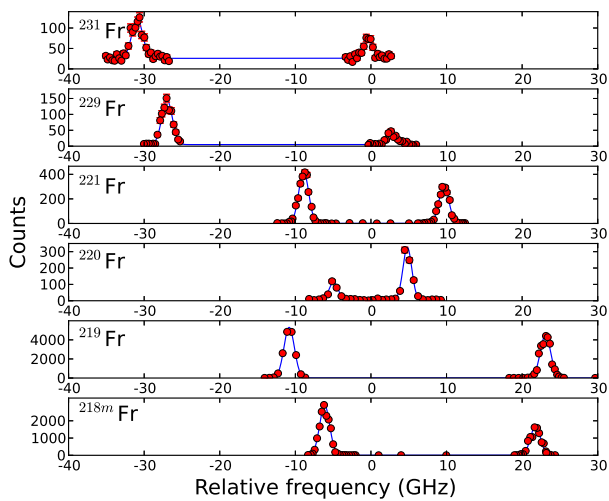}% Here is how to import EPS art

\caption{\label{fig:neutron_rich_hfs}  (Color on-line) Fitted hyperfine spectra for the newly measured $^{218m,219,229,231}\text{Fr}$ isotopes and reference isotopes $^{220,221}\text{Fr}$ .}
\end{figure}

\begin{table}[b]%The best place to locate the table environment is directly after its first reference in text
\caption{\label{tab:moments}%
Extracted hyperfine parameters $A(7s~^{2}S_{1/2})$ and magnetic dipole moments $\mu$, along with spins from literature.
}

\begin{ruledtabular}
\begin{tabular}{clcc}
\textrm{Isotope}&
\textrm{$I^{\pi}$}&
\multicolumn{1}{c}{\textrm{$A(7s~^{2}S_{1/2})$ (GHz)}}&
\textrm{$\mu_{\text{exp}}(\mu_{N})$}\\ 
\colrule \\ [-1.5ex]
\multirow{2}{*}{$^{218m}\text{Fr}$} & $(8^{-})$ \cite{Sheline1999} & $+3.30(3)$ & $+2.68(4)$\\ 
 & $(9^{-})$ \cite{Ewan1982,Debray2000} & $+2.95(3)$ & $+2.70(4)$\\  [1ex]
$^{219}\text{Fr}$ & $9/2^{-}$ \cite{Browne2001} & $+6.82(3)$ & $+3.11(4)$ \\ [1ex]
$^{229}\text{Fr}$ & $(1/2^{+})$ \cite{Borge1992} & $+30.08(11)$ & $+1.53(2)$\\ [1ex]
$^{231}\text{Fr}$ & $(1/2^{+})$ \cite{Fraile2001} & $+30.77(13)$ & $+1.56(2)$\\

\end{tabular}
\end{ruledtabular}
\end{table}
The results for $A(7s~^{2}S_{1/2})$ and $\mu$ for the spins proposed in the literature are given in Table \ref{tab:moments}. In Fig. \ref{fig:neutron_rich_hfs}, the fitted hyperfine spectra are shown for the newly measured $^{218m,219,229,231}\text{Fr}$ isotopes, as well as for the reference isotopes $^{220,221}\text{Fr}$. A minimum error of $30~\text{MHz}$ for $A(7s~^{2}S_{1/2})$ originates from the scatter observed in 18 hyperfine spectra measured throughout the experimental run for the reference isotope $^{221}\text{Fr}$, as shown in Fig. \ref{fig:A_scatter}. For the full discussion on this error, the reader is referred to Ref. \cite{Coco2013,Lynch:1606787}. The largest contribution to the error on the extracted magnetic moments is introduced by the error on the reference magnetic moment of $^{210}\text{Fr}$.

The isotope shifts $\delta\nu^{A,A^{\text{ref}}}$ were determined from the fitted centre-of-gravity of the hyperfine spectra relative to $^{221}\text{Fr}$. For the isotope shift, a minimum error of $100~\text{MHz}$ was established based on the long-term drift and short-term fluctuations of the centroid frequency of $^{221}\text{Fr}$ (a full discussion on this error is available in Ref. \cite{Lynch:1606787}). The change in  nuclear mean-square charge radii can be extracted from the isotope shifts, via

\begin{equation}
\delta\langle r^{2}\rangle^{A,A^{\text{ref}}}=\frac{\delta\nu^{A,A^{\text{ref}}}-\frac{A-A^{\text{ref}}}{A\cdot A^{\text{ref}}}\cdot(K_{NMS}+K_{SMS})}{F},
\label{eq:charge_radii}
\end{equation}

\noindent where $F$ is the field shift, $K_{NMS}$ the normal mass shift and $K_{SMS}$ the specific mass shift constant \cite{Dzuba2005}. The atomic masses $A$ and $A^{\text{ref}}$ are taken from \cite{Wang2012}. The field and mass shift constants depend on the optical transition and in the case of francium have to be calculated theoretically, as in Ref. \cite{Dzuba2005} for the francium $D_{2}$ line (7s~$^{2}S_{1/2}$-7p~$^{2}P_{3/2})$. These constants for the $D_{2}'$ line (7s~$^{2}S_{1/2}$-8p~$^{2}P_{3/2})$, studied in this experiment, could then be determined relative to the calculated values via a King plot \cite{King1984}. The details of this analysis can be found in \cite{LynchNDef2013}. 
 The values extracted using the King-plot method and the formula $K_{NMS}=\nu_{\text{exp}}/1822.888$ \cite{Dzuba2005} were $F_{D2'}=-20670(210)~\text{MHz/fm}^{2}$, $K^{D2'}_{SMS}=360(330)~\text{GHz u}$, and $K^{D2'}_{NMS}=389~\text{GHz u}$. Substituting these values into Eq. \ref{eq:charge_radii} yielded the $\delta\langle r^{2}\rangle^{A,221}$ values given in Table \ref{tab:charge_radii} along with their isotope shifts and assumed spins. The charge radii of $^{229,231}\text{Fr}$ were calculated assuming spins of 1/2 (see Section \ref{sec:mag_moments}). For $^{218m}\text{Fr}$ a spin of 9 is preferred, based on an in-beam spectroscopy study feeding levels in $^{218}\text{Fr}$ that decay to this low lying isomer \cite{Debray2000}, which is suggested to be the fully aligned member of the $\pi(h_{9/2})\otimes\nu(g_{9/2})$ multiplet. However, determining the charge radius using the other spin option of 8 \cite{Sheline1999}, does not lead to a significant different value within error bars. The errors given in Table \ref{tab:charge_radii} for $\delta\langle r^{2}\rangle^{A,221}$ are the uncertainties originating from the experimental isotope shift given in parentheses and the total error taking into account the theoretical errors for the field and mass shift values \cite{Dzuba2005}, given in curly braces.
 
 %Both the isotope shifts and relative mean-square charge radii are given with respect to our measured $^{221}\text{Fr}$ from Coc et al. \cite{Coc1985}, for easy comparison with the radium $\delta\langle r^{2}\rangle^{A,A^{\text{ref}}}$ values in Section \ref{sec:charge_radii}. 

\begin{table}[b]%The best place to locate the table environment is directly after its first reference in text
\caption{\label{tab:charge_radii}%
Extracted isotope shifts $\delta\nu^{A,221}$ and changes in mean-square charge radii $\delta\langle r^{2}\rangle^{A,221}$, along with spins from literature. For $\delta\langle r^{2}\rangle^{A,221}$, the values in parentheses are the experimental uncertainties in determining the isotope shifts, while the values given in curly braces represent the total error taking into account the theoretical errors for the field and mass shift values \cite{Dzuba2005}.
}
\begin{ruledtabular}
\begin{tabular}{clcc}
\textrm{Isotope}&
\textrm{$I^{\pi}$}&
\multicolumn{1}{c}{\textrm{$\delta\nu^{A,221}$ (GHz)}}&
\textrm{$\delta\langle r^{2}\rangle^{A,221}$ (fm$^{2}$)}\\ 
\colrule \\ [-1.5ex]
$^{218m}\text{Fr}$ & $(9^{-})$ \cite{Sheline1999} & +8.24(10) & -0.401(5)\{6\}\\ [1ex]
$^{219}\text{Fr}$ & $9/2^{-}$ \cite{Browne2001} & +5.59(10) & -0.272(5)\{6\}\\ [1ex]
$^{229}\text{Fr}$ & $(1/2^{+})$ \cite{Borge1992} & -18.36(10) & +0.894(5)\{11\}\\  [1ex]
$^{231}\text{Fr}$ & $(1/2^{+})$ \cite{Fraile2001} & -22.14(10) & +1.078(5)\{12\}\\
\end{tabular}
\end{ruledtabular}
\end{table}

\section{\label{sec:Discussion}Discussion}

\subsection{Determination of the state in $^{218m}\text{Fr}$}

For $^{218}\text{Fr}$ it is important to determine whether the measured hyperfine spectrum originates from the isomeric state or from the ground state. The isomeric state has a lifetime of $22.0(5)~\text{ms}$ \cite{Ewan1982}, while the ground state has a lifetime of $1.0(6)~\text{ms}$ \cite{Jain2006}. Since the ion trapping time in ISCOOL was $32~\text{ms}$ in this experimental run, almost all of the $^{218}\text{Fr}$ ground state ions decayed before reaching the laser-ion interaction region. However, the release of the ion bunch from ISCOOL was not synchronized with the proton pulse. This means it is still possible that some $^{218}\text{Fr}$ ground state ions reached the trap if a proton pulse arrived within 5~ms of the trap release trigger.

\begin{figure}
\def\svgwidth{\columnwidth}
\vspace{0.35cm}
\includegraphics[width=\linewidth]{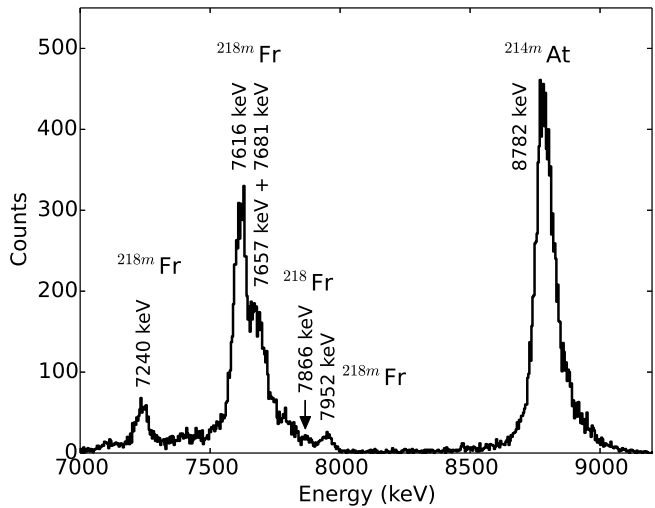}% Here is how to import EPS art
\caption{\label{fig:218_alpha} $\alpha$-particle energy spectrum of $^{218m}\text{Fr}$. The most intense $\alpha$-decay branches of $^{218m}\text{Fr}$ and $^{214m}\text{At}$  observed by Ewan et al. and Sheline et al. \cite{Ewan1982,Sheline1999} are labelled. The vertical arrow indicates the energy of the most intense branch of the ground state $^{218}\text{Fr}$ \cite{Sheline1999}. }
\end{figure}

 Figure~\ref{fig:218_alpha} shows the $\alpha$-decay energy spectrum measured with the DSS upon implanting the non-neutralized component of the $^{218}\text{Fr}$ beam into a carbon foil, with a collection time of 5 minutes. The most intense $\alpha$-decay branches observed by Ewan et al. and Sheline et al. \cite{Ewan1982,Sheline1999} (Table 1 and 2 in \cite{Ewan1982} and Fig.~4a,4b and 4d in \cite{Sheline1999}) are observed here: $7240~\text{keV}, 7616~\text{keV}, 7657~\text{keV} \text,~ 7681~\text{keV},~\text{and}~7952~\text{keV}$ originating from $^{218m}\text{Fr}$ and $8782~\text{keV}$ originating from $^{214m}\text{At}$. The vertical arrow in Fig. \ref{fig:218_alpha} marks the location of the most intense $\alpha$-decay branch of the $^{218}\text{Fr}$ ground state $7866~\text{keV} (I^{\alpha} = 92 \%)$ \cite{Sheline1999}.

\begin{figure}
\def\svgwidth{\columnwidth}
\vspace{0.5cm}
\includegraphics[width=\columnwidth]{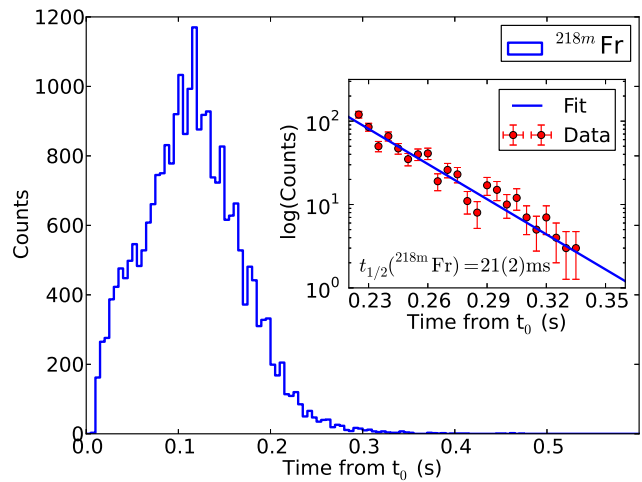}% Here is how to import EPS art
\caption{\label{fig:218_decay}  (Color on-line) Saturation/decay curve of the full $^{218}\text{Fr}$ beam. The x axis shows the time stamp of the detected $\alpha$-particles after the proton impact on target $t_{0}$. The inset shows the fit of the half-life on the exponential of the decay curve.}
\end{figure}

A dedicated half-life measurement was not made in this experimental run for $^{218}\text{Fr}$, however the $\alpha$-decay event timestamp information could be used to estimate the half-life. The individual timestamps were used to create a saturation/decay curve for the full $^{218}\text{Fr}$ beam, shown in Fig.~\ref{fig:218_decay}. The spectrum was obtained by setting the time of the proton impact on the ISOLDE target as $t_{0}$ and plotting the time taken for an $\alpha$-particle to be detected. A $6500-9500~\text{keV}$ energy gate was applied to the spectrum to only include $\alpha$-particles originating from $^{218m}\text{Fr}$ and $^{214m}\text{At}$. Since the half-life of $^{214m}\text{At}$ is $760(15)~\text{ns}$ \cite{Ewan1982}, its decay can be considered virtually instantaneous, and thus the time of detection of a $^{214m}\text{At}$ $\alpha$-particle solely depends on the half-life of its parent $^{218m}\text{Fr}$. Fitting the decaying part of the curve in Fig. \ref{fig:218_decay}, yielded the half-life value: $t_{1/2}(^{218m}\text{Fr}) = 21(2)~\text{ms}$. The value $\chi^{2}_{\text{red}} = 1.59$ from the fit was used to scale up the final uncertainty. The extracted half-life value is in agreement with literature \cite{Ewan1982}, confirming that the measured hyperfine spectrum belongs to the isomeric state $^{218m}\text{Fr}$.

% It should be noted that after implantation of the $^{218}\text{Fr}$ beam, there was not cut-off for the beam coming from the target. This means that there could have still been feeding of the $^{218}\text{Fr}$ ground state via $\alpha$-decay from $^{222}\text{Ac}$, also produced in the target. This could have influenced our value obtained for $t_{1/2}(^{218m}\text{Fr})$. Therefore the value presented here is not to be considered as a new measurement of the half-life of $^{218m}\text{Fr}$, but rather as a firm indication that the hyperfine spectra we obtained at mass $A = 218$ most likely belongs to this isomeric state of $^{218m}\text{Fr}$, and not the ground state.

%\begin{figure}
%\def\svgwidth{\columnwidth}
%\vspace{0.62cm}
%\includegraphics[width=\linewidth]{218Fr_half-life.eps}% Here is how to import EPS art
%\input{218_half_life.eps_tex}
%\caption{\label{fig:218_half_life}  (Color on-line) Determination of the half-life of $^{218m}\text{Fr}$.}
%\end{figure}

\subsection{\label{sec:mag_moments}Magnetic dipole moments}
\begin{figure}
\def\svgwidth{\columnwidth}
%\vspace{-0.3cm}
%\includegraphics[scale=0.40]{test6.eps}% Here is how to import EPS art
\includegraphics[scale=0.45]{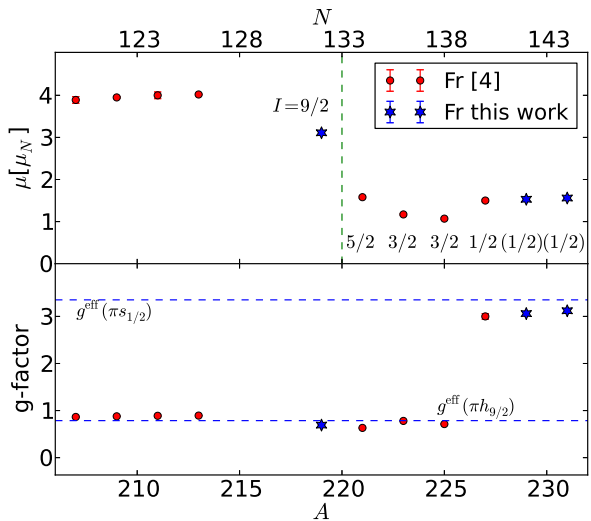}
\caption{\label{fig:mu_odd}  (Color on-line) Magnetic dipole moments (upper panel) along with nuclear spins, and g-factors (lower panel) for $^{219,229,231}\text{Fr}$, together with literature values for the neighbouring odd-even francium isotopes from Ref. \cite{Coc1985}. The $g^{\text{eff}}(s_{1/2})$ values were calculated using $g_{s}^{\text{eff}}= 0.6~g_{s}^{\text{free}}$ and $g_{l} = g_{l}^{\text{free}}$. The vertical dashed line (upper panel) emphasizes where the spin changes from $I~=~9/2$ to the sequence 5/2, 3/2, 3/2, 1/2, (1/2), (1/2). }
\end{figure}

The extracted experimental magnetic dipole moments and g-factors are shown in Fig.~\ref{fig:mu_odd}. The ground-state spin of the odd-$A$ francium isotopes changes with increasing neutron number from $I(^{219}\text{Fr})=9/2$ \cite{Browne2001} to $I(^{227}\text{Fr})=1/2$ \cite{Coc1985}. This is reflected in  the magnetic dipole moment values in Fig.~\ref{fig:mu_odd} (upper panel), while the g-factors (lower panel) remain constant near an effective single particle value. The effective-proton single-particle g-factors for the $\pi 1h_{9/2}$ and $\pi 3s_{1/2}$ orbits are shown, calculated with $g_{s}^{\text{eff}}= 0.6~g_{s}^{\text{free}}$ (typical for this region \cite{Ekstrom1986}) and $g_{l} = g_{l}^{\text{free}}$. These effective values are consistent with numerically calculated single particle magnetic dipole moment values of odd-mass nuclei around $^{208}\text{Pb}$, taking into account mesonic and renormalization corrections as well as core polarization effects \cite{Bauer1973}. Our experimental $g(^{219}\text{Fr}) = +0.69(1)$ value indicates that the $^{219}\text{Fr}$ ground-state wave-function is dominated by an unpaired proton in the $\pi 1h_{9/2}$ orbital, as is the case for the odd-$A$ francium isotopes up to $^{225}\text{Fr}$. The g-factors of these neutron-rich isotopes are systematically lower than those of the neutron-deficient $^{207,209,211,213}\text{Fr}$ isotopes. The latter isotopes, near magic $N~=~126$, have a nearly spherical ground state \cite{Ekstrom1986}, while the neutron-rich isotopes are known to have a deformation larger than $\epsilon_{2} > 0.15$ \cite{Ekstrom1986} (where $\epsilon_{2}$ is the Nillson model quadrupole deformation parameter) and exhibit parity doublet bands \cite{Liang1991} associated with octupole deformations. These deformations have however only little impact on the g-factors because the Nilsson levels are straight lines (they do not mix with other shell model levels).

 From $^{227}\text{Fr}$ onwards the structure of the ground state changes. The spin of $^{227}\text{Fr}$ was measured to be $I~=~1/2$ using the RF-resonance technique in combination with collinear laser spectroscopy \cite{Coc1985}. The g-factor of this state, extracted from the measured magnetic moment $\mu(^{227}\text{Fr}) = 1.50(3)~\mu_{N}$ \cite{Coc1985}, is in good agreement with the effective g-factor for a proton hole in the $\pi 3s_{1/2}$ orbit (Fig.~\ref{fig:mu_odd}).  This suggests that the ground-state wave-function of $^{227}\text{Fr}$ is dominated by a proton-intruder configuration. Based on core-quasiparticle calculations including only quadrupole and hexadecapole deformations \cite{Ekstrom1986}, the ground states of both $^{227}\text{Fr}$ and $^{228}\text{Fr}$ are found to be dominated by a proton hole in the $\pi 3s_{1/2}$ orbital (1/2[400] orbit). The spin and magnetic dipole moment for $^{227}\text{Fr}$  could also be reproduced by the reflection-asymmetric rotor model \cite{Leander1988},  with 1/2[400] as the dominant component of the wave-function, assuming that the octupole deformation parameter is $\beta_{3}~=~0$ (where $\beta_{3}$ is the octupole deformation parameter expressed in the spherical harmonics expansion). The main factor determining the level ordering in the reflection-asymmetric rotor model is the quadrupole deformation parameter, which increases for francium with increasing neutron number \cite{Coc1987}. The trend of increasing values of the quadrupole moment and $\beta_{2}$-deformation parameter with increasing neutron number has also been observed in the thorium ($Z~=~90$) and uranium ($Z~=~92$) isotopes \cite{Bemis1973}. The ground-state spin of $^{229}\text{Fr}$ was also tentatively assigned to be $I~=~(1/2)$ \cite{Borge1992}, arguing that it should be the same as for $^{227}\text{Fr}$ unless the quadrupole deformation of $^{229}\text{Fr}$ would be significantly smaller than $^{227}\text{Fr}$ and the octupole deformation significantly larger, both being unlikely. The spin of $^{231}\text{Fr}$ is tentatively assigned as (1/2) based on a characteristic $\beta-\text{decay}$ pattern between the 1/2[400] and 1/2[501] Nilsson configurations observed in the vicinity of $^{231}\text{Ra}$ \cite{Fraile2001}.
 
\begin{figure}
\def\svgwidth{\columnwidth}
\vspace{+0.67cm}
\includegraphics[width=\columnwidth]{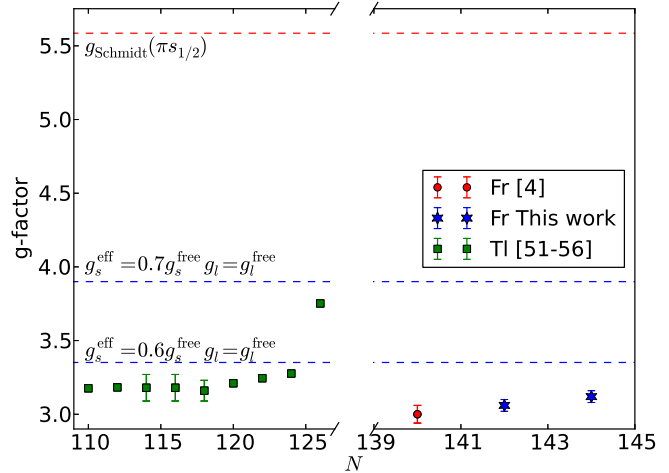}% Here is how to import EPS art
%\input{Tl_eps2_gray.eps_tex}
%\resizebox{\linewidth}{7.8cm}{\input{Tl_eps2_gray.eps_tex}}
%\input{Tl_eps2_gray.eps_tex}
\caption{\label{fig:Tl} (Color on-line)  Comparison of $^{227,229,231}\text{Fr}$ and odd-$A$ $^{191-207}$Tl g-factors. The data for francium were taken from Ref. \cite{Coc1985} and this work, while the thallium data is from Refs. \cite{Menges1992,Bounds1987,Bengtsson1984,Ekstrom1975,Raghavan1989,Neugart1985}. The $g^{\text{eff}}(s_{1/2})$ values were calculated using $g_{s}^{\text{eff}}~=~0.6~g_{s}^{\text{free}}, g_{l} = g_{l}^\text{free}$ and $g_{s}^{\text{eff}}= 0.7~g_{s}^{\text{free}}, g_{l} = g_{l}^{\text{free}}$. }
\end{figure}

\begin{figure}
\includegraphics[scale=0.45]{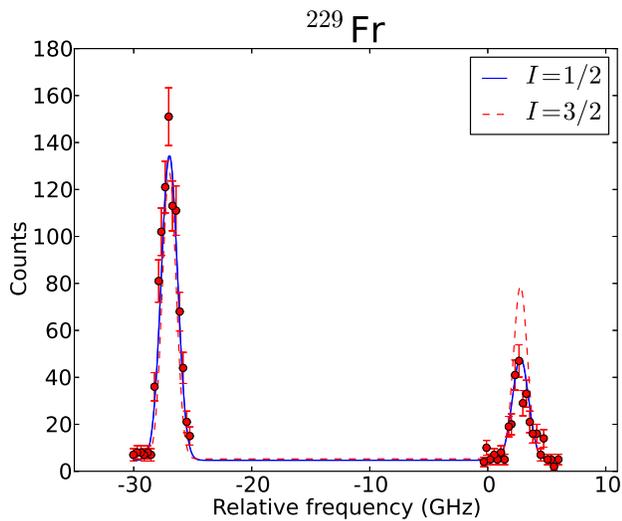}% Here is how to import EPS art
\caption{\label{fig:int_ratio} (Color on-line)  Fits to experimental data assuming the values $I~=~(1/2)$ and $I~=~(3/2)$ for $^{229}\text{Fr}$. The better agreement with experiment for $I~=~(1/2)$ is evident.}
\end{figure}

 Our experimental $g(^{229}\text{Fr}) = +3.06(4)$ and $g(^{231}\text{Fr}) = +3.12(4)$ values (Fig.~\ref{fig:mu_odd}) are indeed consistent with a proton hole in the $\pi 3s_{1/2}$ orbital (the 1/2[400] orbit), as was proposed in \cite{Ekstrom1986} for $^{227,228}\text{Fr}$. The $^{227,229,231}\text{Fr}$ g-factors also agree with the g-factors  of the odd Tl isotopes, which have a hole in the $\pi 3 s_{1/2}$ orbit (illustrated in Fig.~\ref{fig:Tl}). On the neutron-rich side ($N~>~140$), the $\pi 3s_{1/2}$ 1/2[400] intruder state has been assigned to the ground states of $^{231}$Ac and $^{237}$Pa \cite{Thompson1977}. Assuming the same behaviour for the increase of quadrupole moment and quadrupole deformation parameter with increasing mass as previously mentioned for thorium and uranium, the authors in Ref. \cite{Thompson1977} comment that as a hole state, the 1/2 [400] excitation energy decreases with increasing $\epsilon_{2}$ and with increasing mass, for the actinium ($Z~=~89$) and protoactinium ($ Z~=~91$) isotopes (Fig.~3 in \cite{Thompson1977}). It is reasonable to assume a similar behaviour of this level in francium, whereby it would follow that this level remains the ground state as more neutrons are added to $^{227}\text{Fr}$. The tentative (1/2) spin assignment for $^{229}\text{Fr}$ fits with the intensity ratios of the HFS peaks in our experimental spectra, as shown in Fig.~\ref{fig:int_ratio}. Due to angular momentum coupling considerations, the intensity ratio between the two groups of transitions from the $7s~^{2}S_{1/2}$ ground state to the $8p~^{2}P_{3/2}$ excited state differs noticeably in the case of $I~=~1/2$ and $I~>~1/2$. A better agreement between the fit and experimental data is observed for $I~=~1/2$ for $^{229}\text{Fr}$. For $^{231}\text{Fr}$ however, this procedure did not lead to conclusive results, due to an increased experimental background (most likely originating from radium) and lower statistics observed in the $^{231}\text{Fr}$ hyperfine spectrum.

For $^{218m}\text{Fr}$, there is no firm spin assignment. Debray et al. \cite{Debray2000} tentatively assigned the spin value $I~=~(9^{-})$ to $^{218m}\text{Fr}$, based on an in-beam spectroscopy study feeding levels in $^{218}\text{Fr}$ that decay to this low lying isomer. This state is suggested to be the fully aligned member of the $\pi(h_{9/2})\otimes\nu(g_{9/2})$ multiplet. The second possible tentative spin-parity assignment is $(8^{-})$, based on $\alpha$-decay feeding of several states of the $\pi(h_{9/2})\otimes\nu(g_{9/2})$ multiplet in $^{214}\text{At}$ \cite{Sheline1999}. With these tentative spins the hyperfine parameters are extracted from the data, and related magnetic moments are determined (Table \ref{tab:moments}). A further discussion on the structure of the isomeric state based on these moments is not possible, because calculated effective and empirical g-factor values for different possible configurations of this odd-odd isomer do not favor one or the other configuration. A firm spin assignment will provide a first step to better understanding the structure of this isomeric state.
%Sheline et al. \cite{Sheline1999} tentatively assigned the spin value $I~=~(8^{-})$ to $^{218m}\text{Fr}$,  ,

\subsection{\label{sec:charge_radii}Mean-square charge radii}
\begin{figure}
\def\svgwidth{\columnwidth}
\vspace{-0.1cm}
\includegraphics[scale=0.43]{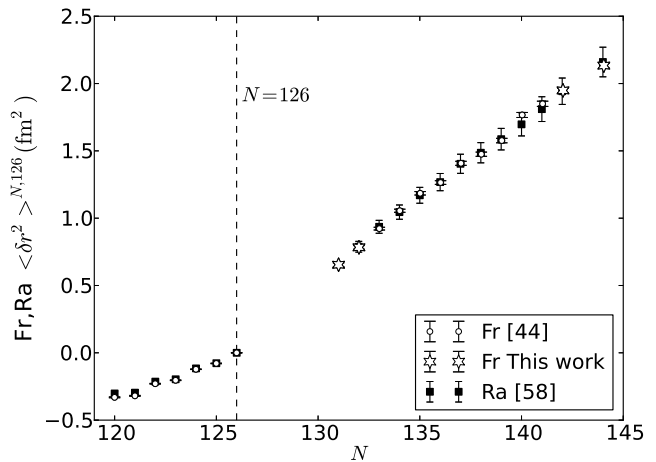}
\caption{\label{fig:charge_radii} Comparison of $\delta\langle r^{2}\rangle^{N,126}$ values for francium and radium. The francium $\delta\langle r^{2}\rangle^{N,126}$ values given in circles were taken from Dzuba et al. \cite{Dzuba2005}. The radium $\delta\langle r^{2}\rangle^{N,126}$ values were taken from Wansbeek et al. \cite{Wansbeek2012}.}
\end{figure}

In order to compare our $^{218m,219,229,231}\text{Fr}$ mean-square charge radii values with the ones from Dzuba et al. for \ce{^{220--228}Fr} \cite{Dzuba2005}, we converted our $\delta\langle r^{2}\rangle^{A,221}$ values and the Dzuba $\delta\langle r^{2}\rangle^{A,212}$ values to be with respect to $^{213}\text{Fr} (N = 126)$. The obtained $\delta\langle r^{2}\rangle^{N,126}$ values for francium can then be compared to the ones for radium \cite{Wansbeek2012}, defined relative to $^{214}_{88}\text{Ra} (N = 126)$. The comparison is shown in Fig. \ref{fig:charge_radii}.
%In order to compare the new $^{218m,219,229,231}\text{Fr}$ relative mean-square charge radii values, along with values from literature for \ce{^{220--228}Fr} \cite{Dzuba2005}, to the ones for radium \cite{Wendt1987,Wansbeek2012}, we converted our measured $\delta\langle r^{2}\rangle^{A,221}$ values and the Dzuba $\delta\langle r^{2}\rangle^{A,212}$ values to be with respect to $^{213}Fr (N = 126)$.
%Figure~\ref{fig:charge_radii} plots the obtained $\delta\langle r^{2}\rangle^{N,126}$ values for francium and radium.
 The neutron-rich radium isotopes have long been associated with octupole deformations (\cite{Butler1996} and references therein) and most noticeably $^{224}\text{Ra}$ displays a stable reflection-asymmetric deformation of its nuclear shape \cite{Gaffney2013}. The theoretical error bands for radium are significant \cite{Wansbeek2012}, making the comparison difficult, but nevertheless Fig.~\ref{fig:charge_radii} shows that the two isotopic chains follow a very similar trend along the whole mass range. The newly measured $^{218m,219,229,231}\text{Fr}$ $\delta\langle r^{2}\rangle^{A,213}$ values follow the trend of the  $\text{Fr}_{133-141}$ isotopes measured by Coc. et al. \cite{Coc1985}.

%$^{218m,219,229,231}\text{Fr}$ obtained in this work together with previous known values from \cite{Coc1985},

%\begin{figure}
%\def\svgwidth{\columnwidth}
%\vspace{1cm}
%%\includegraphics[scale=0.40]{charge_radii_ink.eps}% Here is how to import EPS art
%\input{charge_radii_zoomed.eps_tex}
%\caption{\label{fig:charge_radii_zoomed} Comparison of $\delta\langle r^{2}\rangle^{N,126}$ values for francium and radium for $N = 131-144$.}
%\end{figure}

The odd-even staggering (OES) effect of the changes in mean-square charge radii has been associated with reflection-asymmetric nuclear shapes \cite{Coc1987,Ahmad1988}. In most nuclei the OES order indicates that nuclei with odd-$N$ have a smaller mean-square charge radius with respect to the average of their even-$N$ neighbours. This is considered as normal OES ordering. The OES effect can be described by the $D(N;\delta\langle r^{2}\rangle^{N,126})$ factor, expressing how the change in mean-square charge radius deviates from the mean of its neighbours:

\begin{eqnarray}
D(N;\delta\langle r^{2}\rangle^{N,126}) &=&(-1)^{N} \biggl[\delta\langle r^{2}\rangle^{N,126}\bigr.
\nonumber  \\
 & & \hspace{-2em} \left.-
 \frac{\delta\langle r^{2}\rangle^{(N-1),126}+\delta\langle r^{2}\rangle^{(N+1),126}}{2}\right].
 \label{eq:D_factor}
\end{eqnarray}

\begin{figure}
\def\svgwidth{\columnwidth}
%\vspace{0.2cm}
%\includegraphics[width=\columnwidth]{staggering_no_219_gray}
\includegraphics[scale=0.45]{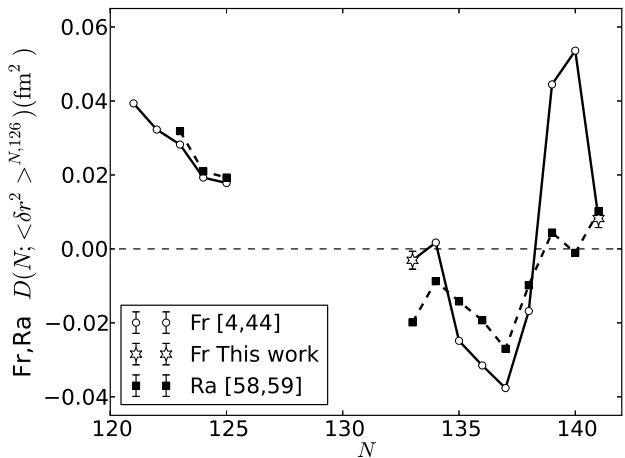}
\caption{\label{fig:odd_even_staggering} Comparison of $D(N;\delta\langle r^{2}\rangle^{N,126})$ values for francium and radium. Only the uncertainties originating from the experimental isotope shift for the $\delta\langle r^{2}\rangle^{N,126}$ values were taken into account \cite{Coc1985,Wendt1987}, due to the large theoretical errors on the $\delta\langle r^{2}\rangle^{N,126}$ values for radium \cite{Wansbeek2012}.  }
\end{figure}

Eq. \ref{eq:D_factor} is defined the same way as in Ref. \cite{Coc1987}, except that $\delta\nu$ was used in place of $\delta\langle r^{2} \rangle$. In the present discussion $D(N;\delta\nu)$ cannot be used due to the different transitions under consideration. Coc. et. al. \cite{Coc1987} attributed normal OES ordering to $D(N;\delta\nu)~<~0$. This type of ordering is seen in most nuclei, as in cesium ($ Z~=~55$) for example \cite{Coc1987}. Since we compare $D(N;\delta\langle r^{2}\rangle^{N,126})$ and the field shift constant has a negative sign, the values of $D(N;\delta\langle r^{2}\rangle^{N,126})$ compared to $D(N;\delta\nu)$ will have an opposite sign and therefore we attribute normal OES ordering to $D(N;\delta\langle r^{2}\rangle^{N,126})~>~0$. Figure~\ref{fig:odd_even_staggering} shows the result of applying Eq. \ref{eq:D_factor} to the radii from Fig.~\ref{fig:charge_radii}, taking only the experimental error from the isotope shift into account \cite{Coc1985,Wendt1987} and neglecting the systematic theoretical error from the uncertainty of the mass and field shift values \cite{Dzuba2005,Wansbeek2012}.

A similar approach was used in Ref. \cite{Ahmad1988} for the study of the radium charge radii, from which it was concluded that between $^{220}\text{Ra}$ and $^{226}\text{Ra}$ the inverted OES points to octupole deformation in the ground state being well developed for the odd-$N$ isotopes ($^{221,223,225}\text{Ra}$). They also offered a qualitative interpretation of the correlation between the inversion in the OES order with reflection-asymmetric shapes, in terms of the different influences of pairing correlations on the shape of the nuclear potential between odd- and even-$N$ nuclei (Fig. 7 in Ref. \cite{Ahmad1988}).

Our $D(N;\delta\langle r^{2}\rangle^{N,126})(^{220}\text{Fr})$ value lies slightly bellow zero, which implies that it sits at the border of the region with reflection-asymmetric shapes (having inverted OES). That is confirmed by earlier studies \cite{Ekstrom1986,Sheline1988}, where observables could be interpreted without including stable octupole deformations, while other observables did include some degree of octupole deformation. Indeed, in Ref. \cite{Ekstrom1986}, the authors could not adequately interpret the configuration of $^{220}\text{Fr}$ using core-quasiparticle calculations without octupole degrees of freedom. Similarly in Ref. \cite{Sheline1988}, the spin and parity $I(^{220}\text{Fr})~=~1^{+}$ could be interpreted by including octupole deformations. Evidence for reflection asymmetry was also found in the form of parity doublet decay bands by \cite{Liang1992}, who interpreted this isotope to be in the transition region between quadrupole-octupole deformations and spherical symmetry.

The positive value for $D(N;\delta\langle r^{2}\rangle^{N,126})(^{228}\text{Fr})$ establishes a normal OES extending from $^{226}\text{Fr}$ onwards. However, the value for $^{228}\text{Fr}$ is considerably lower than the $D(N;\delta\langle r^{2}\rangle^{N,126}(^{226,227}\text{Fr}))$ values. The $\mu(^{228}\text{Fr})$ and $Q_{2}^{s}(^{228}\text{Fr})$ values were qualitatively well reproduced by Ref. \cite{Ekstrom1986} without including octupole deformations, while in Ref. \cite{Sheline1988} the $I(^{228}\text{Fr})=2^{-}$ value is reproduced only by taking octupole deformation into account. Our positive $D(N;\delta\langle r^{2}\rangle^{N,126})(^{228}\text{Fr})$ value supports the absence of octupole deformations in the ground state of $^{228}\text{Fr}$.

%Since $^{219}\text{Fr}$ exhibits parity-doublet decay bands \cite{Liang1991} and is believed to possess reflection-asymmetric deformation of its nuclear shape \cite{Sheline2000}, then it would seem this isomeric state also exhibits this type of deformation.

\section{Summary and Conclusions}

The hyperfine spectra of $^{218m,219,229,231}\text{Fr}$ have been measured using the new Collinear Resonance Ionization Spectroscopy (CRIS) beam line at ISOLDE, CERN. From the measured spectra, the magnetic dipole moment, and changes in mean-square charge radii values  were extracted based on known and assumed spin values. This allowed nuclear structure conclusions to be drawn for these nuclei. The isomeric state $^{218m}\text{Fr}$ and the ground state of $^{219}\text{Fr}$ seem to possess the same degree of deformation, based on their mean-square charge radii. For $^{219}\text{Fr}$, its spin-parity $9/2^{-}$ and $g(^{219}\text{Fr}) = +0.69(1)$ value are consistent with the unpaired valence proton occupying the $ 1h_{9/2}$ orbital. The $\mu(^{219}\text{Fr})$ value is smaller than the more neutron deficient $^{207,209,211,213}\text{Fr}$ isotopes, which are well described by the shell model $\pi 1h_{9/2}$ state and considered weakly deformed \cite{Ekstrom1986}. Its g-factor agrees well with those of $^{221,223,225}\text{Fr}$ isotopes, known to be well-deformed with $\epsilon_{2} > 0.15$ \cite{Ekstrom1986}. The small negative value for $D(N;\delta\nu)(^{220}\text{Fr})$, supports the interpretation of this isotope being at the border of the region dominated by octupole deformations \cite{Liang1992}. For $^{228}\text{Fr}$ there is no clear consensus on the presence of reflection asymmetry, but the relatively large positive  $D(N;\delta\langle r^{2}\rangle^{N,126})(^{228}\text{Fr})$ value implies this nucleus lies outside of the region of reflection asymmetry. The $g(^{229}\text{Fr}) = +3.06(4)$ and $g(^{231}\text{Fr}) = +3.12(4)$ values, agree with the unpaired valence proton occupying the $\pi 3s_{1/2}$ intruder orbital. These values also compare well to the g-factor values of odd-$A$ thallium isotopes, having a single $\pi 3s^{-1}_{1/2}$ hole state in the $\text{Z}~=~82$ shell. For $^{229}\text{Fr}$, the relative intensity ratios between the hyperfine structure peaks favor an $I=(1/2)$ spin assignment. Even though these considerations cannot be used to confirm the nuclear spins for $^{229,231}\text{Fr}$, they do support the previous tentative $I^{\pi}(^{229,231}\text{Fr})~=~(1/2^{+})$ \cite{Borge1992,Fraile2001} spin assignments. New theoretical work is required to better understand the specific structure of the neutron-rich francium isotopes. Further experimental information will be obtained using a narrow linewidth laser system, enabling the measurement of spin and spectroscopic quadrupole moments ($I~>~1/2$), which will provide further information on the deformations of these nuclei. 

\section*{Acknowledgements}

We would like to thank the ISOLDE technical group for their support and assistance, and the GSI Target Laboratory for providing the carbon foils. This work has been supported by  the BriX Research Program No.~P7/12 and FWO-Vlaanderen (Belgium)  and GOA 10/010 from KU Leuven, the Science and Technology Facilities Council consolidated grant ST/F012071/1 and continuation grant ST/J000159/1, and the EU Seventh Framework through ENSAR(506065). K.T. Flanagan was supported by STFC Advanced Fellowship Scheme grant number ST/G006415/1.  We acknowledge the financial aid from the Ed Schneiderman Fund at NYU.

\bibliography{neutron_rich_paper}% Produces the bibliography via BibTeX.

\end{document}